\documentclass{jfm}
\usepackage{epstopdf, epsfig}

\usepackage{graphicx}% Include figure files
\usepackage{tabularx}
\usepackage{graphicx}% Include figure files
\usepackage{bm}% bold math
\usepackage{array}
\usepackage{makecell}
\usepackage{natbib}
\usepackage{physics}
\usepackage{float}
\usepackage{todo}
\usepackage{enumitem}
\usepackage{multirow}
\usepackage{tabularx}
\usepackage{gensymb}
\usepackage{marginnote}
\usepackage{amsmath}
\usepackage{enumerate}
\usepackage{mathrsfs}

\newcommand{\imagi}{\mathrm{i}}

\newcommand{\epsp}{\epsilon_p}
\newcommand{\hin}{h_{\text{IN}}}
\newcommand{\Qm}{\text{Q}_m}
\newcommand{\Ds}{\text{D}_s}
\newcommand{\OD}{\text{OD}}

\newcommand{\red}[1]{\ifmmode\mathbf{\textcolor{red}{#1}}\else \textbf{\textcolor{red}{#1}}\fi}
\usepackage{xcolor}

\shorttitle{Fibre coating dynamics influenced by nozzle geometry}
\shortauthor{H. Ji, A. Sadeghpour, Y. S. Ju, A. L. Bertozzi}
\title{Modeling film flows down a fibre influenced by nozzle geometry}% Force line breaks with \\
\author{H. Ji\aff{1}
  \corresp{ \email{hangjie@math.ucla.edu}}{$\ddagger$},
    A. Sadeghpour\aff{2}
     \corresp{These authors contributed equally to this paper.},
  Y. S. Ju\aff{2}, \and A. L. Bertozzi\aff{1,2}}

\affiliation{\aff{1}Department of Mathematics, University of California, Los Angeles
Los Angeles, CA 90095, USA
\aff{2}Mechanical and Aerospace Engineering Department, University of California, Los Angeles
Los Angeles, CA 90095, USA}

\date{\today}% It is always \today, today,
\begin{document}

\maketitle 

\begin{abstract}

We study the effects of nozzle geometry on the dynamics of thin fluid films flowing down a vertical cylindrical fibre. Recent experiments show that varying the nozzle diameter can lead to different flow regimes and droplet characteristics in the film. Using a weighted residual modeling approach, we develop a system of coupled equations that account for inertia, surface tension effects, gravity, and a film stabilization mechanism to describe both near-nozzle fluid structures and downstream bead dynamics. We report good agreement between the predicted droplet properties and the experimental data.
\end{abstract}

\begin{keywords}
%Authors should not enter keywords on the manuscript, as these must be chosen by the author during the online submission process and will then be added during the typesetting process.
\end{keywords}

%\end{CJK*}

%\tableofcontents %not compatible 

\section{Introduction}
The dynamics of thin fluid films flowing down a cylindrical fibre plays a significant role in a variety of engineering applications, including mass and heat exchangers for thermal desalination and water vapor and CO${_2}$ capture (\cite{sadeghpour2019water, zeng2019highly}$)$.  
These films exhibit complex interfacial flow dynamics. Three distinct flow regimes are observed in previous experiments by \cite{kliakhandler2001viscous}: $(a)$ a convective regime where irregular wave patterns frequently lead to droplet collisions; $(b)$ a Rayleigh-Plateau regime where stable traveling beads move at a constant speed, and $(c)$ an isolated droplet regime where widely spaced traveling beads coexist with secondary small-amplitude wave patterns. 
These dynamic regimes have been extensively studied both experimentally and theoretically (\cite{quere1999fluid,kalliadasis2011falling,ruyer2012wavy,ruyer2009film}) as a function of the flow rate and fibre radius for different fluids. 

%previous models
Classical lubrication theory is widely applied to study the dynamics of films flowing down vertical fibres at small flow rates. Under the assumption that the film thickness is much smaller than the fibre radius, weakly nonlinear thin-film equations are investigated in the work of \cite{frenkel1992nonlinear,chang1999mechanism,kalliadasis1994drop}. These evolution equations capture both stabilizing and destabilizing roles of the surface tension that origin from axial and azimuthal curvatures of the interface, respectively (\cite{craster2009dynamics}). A fully nonlinear curvature term was incorporated in \cite{kliakhandler2001viscous} to alleviate limitations of the small-interface-slope assumption of the lubrication theory.
Using a low-Bond-number, surface-tension-dominated theory, \cite{craster2006viscous} propose an asymptotic (CM) model that captures the flow regimes (a) and (c).

Recently, \cite{ji2019dynamics} investigate a full lubrication model that includes slip boundary conditions, nonlinear curvature terms, and a film stabilization term. The last term brings to focus the presence of a stable liquid layer that plays an important role in the dynamics. Compared with models from previous studies, the combination of these physical effects better characterizes the observed propagation speed, stability of traveling droplets, and their transition to the isolated droplet regime. 
For moderate flow rate cases,  \cite{trifonov1992steady} proposed a system of coupled evolution equations for the film thickness and the flow rate. This model incorporates inertial effects based on the integral boundary layer (IBL) equations for the dynamics of a falling film on inclined planes (\cite{shkadov1967wave}). 
Weighted residual integral boundary-layer (WRIBL) models developed by \cite{ruyer2008modelling, duprat2009spatial,ruyer2012wavy} further extended the IBL models by including the effects of the streamwise viscous diffusion. 

These previous studies, however, primarily address the dynamics of downstream flows far away from the inlets.  A recent experimental study by \cite{sadeghpour2017effects} reveals that the geometry of the inlet nozzle also has a strong influence on the downstream dynamics. Specifically, distinct regimes of interfacial patterns are observed by simply varying the diameter of the nozzle while keeping other parameters fixed.
These results motivate us to further investigate the existing models to better understand the relevant influential physics both near the nozzle and further downstream. An improved understanding of the flow regimes will provide insights for a variety of engineering applications.
In this study, we build on previous studies of IBL equations and the film stabilization mechanism, while accounting for inertial effects, gravity modulation, and surface tension.  We compare a new model to experiments with varying nozzle geometry.

The paper is organized as follows. In section \ref{sec:exp} we lay out the experimental setup. In section \ref{sec:model} a system of coupled evolution equations for the film thickness and the flow rate is formulated. Section \ref{sec:stability} presents the stability analysis of the model and discusses the film stabilization term. Numerical results for the model and their comparison to experimental observations are presented in section \ref{sec:numerics} and section \ref{sec:beads}, followed by concluding remarks in section \ref{sec:conclusion}.

\section{Experiments}
\label{sec:exp}
\begin{figure}
    \centering
\includegraphics[width = 6cm]{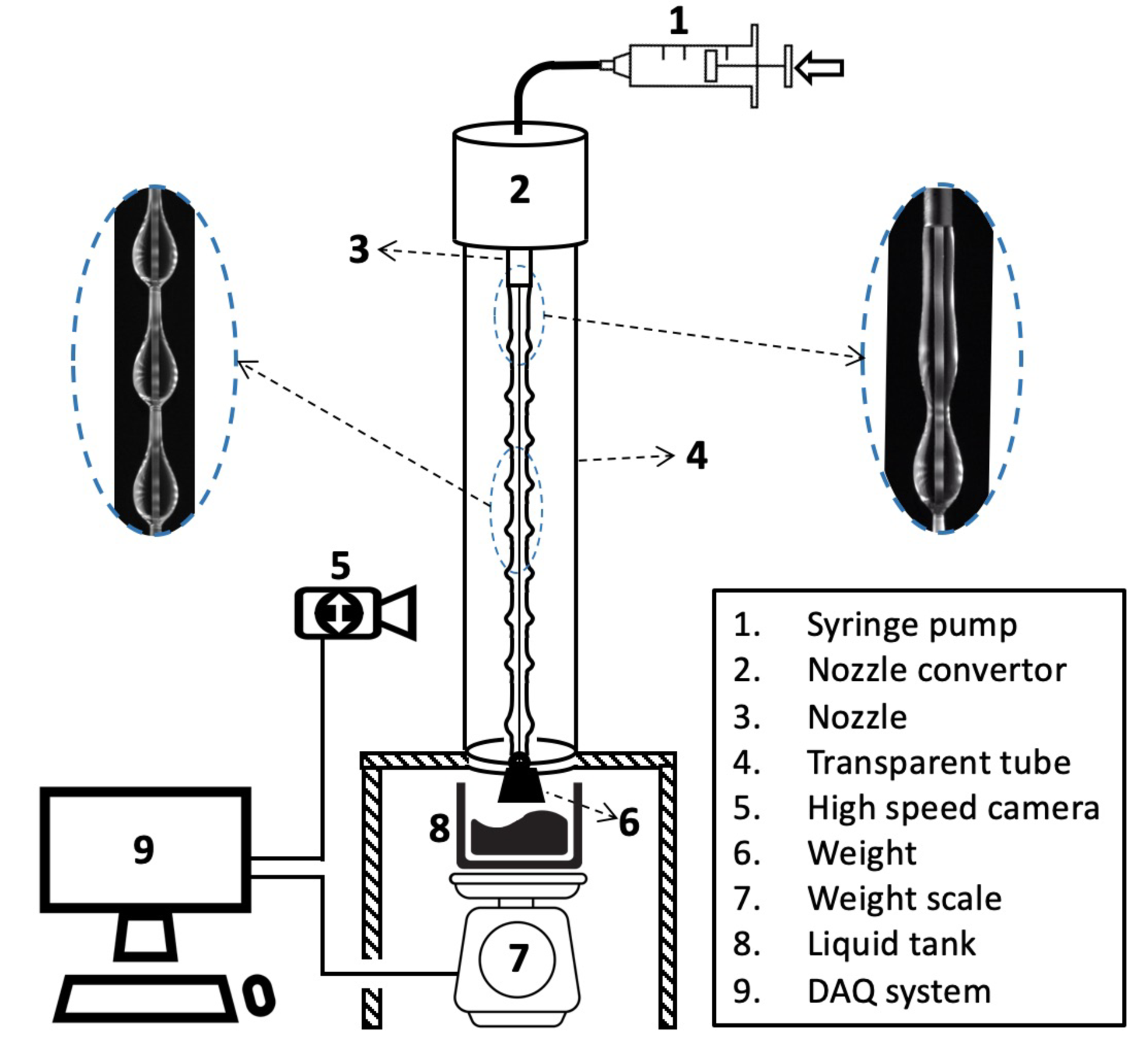}  \hspace{0.5in}  
\includegraphics[width = 2cm]{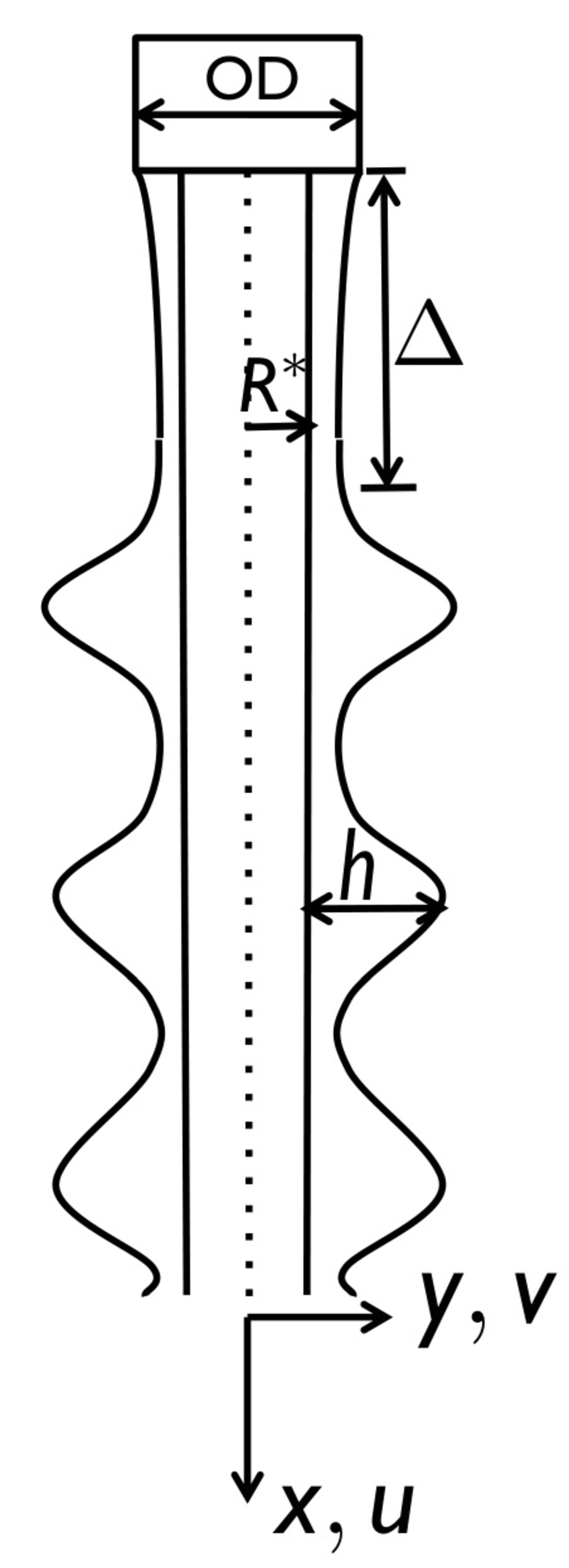}
    \caption{Schematic of the experimental setup with changeable nozzle to study the effect of nozzle diameter (\OD) on the fluid dynamics of the flow}
    \label{fig:experimental_setup}
\end{figure}
Figure \ref{fig:experimental_setup} shows a schematic of the experimental setup, designed to investigate the effects of the inlet nozzle on the flow properties and flow pattern. The experimental setup includes: (1) a syringe pump to control the volume flow rate of the working liquid, (2) a converter to connect the syringe pump outlet to the nozzle, (3) a stainless steel nozzle with various diameters ($\OD$ = $0.84$, $1.06$, $1.27$, $1.56$, $1.86$, or $2.41$ mm), (4) a transparent tube to protect the flow from noise and air disturbances, (5) a high-speed camera, set to $1000$ frame/second, mounted on an adjustable stage, (6) a weight connected to the end of the polymer fibre to keep it straight and vertical during the experiment, (7) a weight scale, (8) a liquid container and (9) a data acquisition system to receive/record the information from the camera and the weight scale.

Polymeric fibres with diameters of $\Ds=0.2$, $0.29$, and $0.43$ mm are coated with a liquid mass flow rate in the range of $\Qm=0.02$ g/s to $0.08$ g/s. The wetting liquid is Rhodorsil silicone oil v50, with the following physical properties: density $\rho=963$ kg/m$^3$, kinematic viscosity $\nu = 50 ~\text{mm}^2$/s, surface tension  $\sigma= 20.8$ mN/m (all at 20\degree C). 
A summary of the experimental conditions is presented in Table \ref{table:experiment}.
\begin{table}
\centering
\begin{tabular} {lcccccc}
Fibre diameter && Nozzle outer diameter  && Mass flow rate  \\
$\Ds$ (mm) && $\OD$ (mm) && $\Qm$ (g/s) \\
 &&&&&&\\
0.2  && 0.84, 1.06, 1.27, 1.56, 1.86, 2.41  && 0.04, 0.08 \\
0.29 && 0.84, 1.27, 1.56, 1.86, 2.41 && 0.04, 0.08 \\
0.43 && 0.84, 1.06, 1.27, 1.56, 1.86, 2.41 &&
0.02, 0.06  \\
&&&&&&\\
 \end{tabular}
  \caption{Experimental cases of varying nozzle diameter and mass flow rate}
  \label{table:experiment}
\end{table}

\section{Model formulation}
\label{sec:model}
We build on the first-order weighted residual integral boundary-layer (WRIBL) model \cite{ruyer2008modelling,duprat2009spatial}. We consider a flow of two-dimensional axisymmetric Newtonian fluid coating a vertical cylinder of radius $R^*$. The kinematic viscosity $\nu$, the density $\rho$ and the surface tension $\sigma$ of the liquid are assumed to be constant.
Near the nozzle, the fluid coats the fibre with a nearly uniform film that defines the characteristic radial lengthscale $\mathcal{H}$. The length of this near-nozzle region, from the inlet to the location of the onset of droplet formation, is referred to as the \emph{healing length} ($\Delta$ in figure \ref{fig:experimental_setup}) (\cite{sadeghpour2017effects}).
Given the dimensional volumetric mass flow rate $\Qm$ and fibre radius $R^*$, we define the volumetric flow rate per circumference unit $q_0^* = \Qm/(2\pi\rho R^*)$. By balancing the viscosity and gravity acceleration, the axial velocity $u_0^*$ of a uniform flow, without interfacial variance in the streamwise direction,
is given by the Nusselt solution
\begin{equation}
u_0^*(y^*) = \frac{g}{\nu}\left[-\frac{1}{4}(y^{*2}-R^{*2})+\frac{1}{2}(\mathcal{H}+R^*)^2\ln\left(\frac{y^*}{R^*}\right)\right].
    \label{eq:u_profile}
\end{equation}
One then obtains the characteristic axial lengthscale $\mathcal{H}$ by solving the equation \eqref{q:nusselt_dim}
\begin{equation}
q_0^* \equiv \frac{1}{R^*}\int_{R^*}^{R^*+\mathcal{H}} u^*_0 y^* ~dy^* =\frac{g}{\nu}\frac{\mathcal{H}^{3}}{3}\phi\left(\frac{\mathcal{H}}{R^*}\right),
    \label{q:nusselt_dim}
\end{equation}
where
$
   \phi(X) = \textstyle{\frac{3}{16X^3}}[(1+X)^4(4\log(1+X)-3)+4(1+X)^2-1].
$

The following scales are introduced for the model: the lengthscale in the streamwise direction $x$ is $\mathcal{L} = \mathcal{H}/\kappa$, where the scale ratio $\kappa = (\rho g \mathcal{H}^2/\sigma)^{1/3}$ is set by the balance between the surface tension and the gravity. 
The characteristic streamwise velocity is $\mathcal{U}=(g\mathcal{H}^2\phi(\alpha))/{\nu}$, where $\alpha = \mathcal{H}/R^*$ represents the ratio of the characteristic film thickness and the fibre radius.
The timescale is $\mathcal{T} = \mathcal{L} /\mathcal{U}$ and the Reynolds number is $\Re = {q_0^*}/\nu$. 

Using these scalings, we follow the analysis of \cite{ruyer2008modelling} and derive a coupled system of two evolution equations for the dimensionless film thickness $h$ and flow rate $q$.
Here the dimensionless flow rate $q$ is defined by 
$q = R^{-1}\textstyle{\int_{R}^{R+h} u y ~dy}$,
where $R$ and $u$ are dimensionless fibre radius and radial velocity. By projecting the velocity field on a set of test functions and applying a weighted residual procedure under the lubrication assumption,
we obtain a mass conservation equation for $h$ in terms of $q$:
\begin{subequations}
\begin{equation}
    \frac{\partial h}{\partial t} = -\frac{1}{1+\alpha h}\frac{\partial q}{\partial x},
    \label{eq:h}
\end{equation}
and an averaged axial momentum equation:
\begin{equation}
\begin{split}
   \delta \frac{\partial q}{\partial t} =  & \delta\left[- F(\alpha h)\frac{q}{h}\frac{\partial q}{\partial x} + G(\alpha h)\frac{q^2}{h^2}\frac{\partial h}{\partial x}\right]\\+
   & \frac{I(\alpha h)}{\phi(\alpha)}\left[-\frac{3\phi(\alpha)}{\phi(\alpha h)}\frac{q}{h^2}+ h\left\{1-\frac{\partial}{\partial x}\left[
   \mathcal{Z}(h) - \frac{\partial^2 h}{\partial x^2}\right]\right\}\right].
   \label{eq:q}
\end{split}
\end{equation}
\label{eq:main}
\end{subequations}
Here $\delta = 3\kappa\Re$ is a reduced Reynolds number, and the coefficients $F, G, I$ are functions of the aspect ratio $\alpha$ and $h$ defined as follows:
\begin{equation}
I(X) = \textstyle{{[64X^5\phi(X)^2]}/{[3F_b(X)}]},\nonumber
\label{eq:phiI}
\end{equation}
\begin{equation}
    F(X) = [{3F_a(X)}]/[{16X^2\phi(X)F_b(X)}],
    \quad G(X) = [{G_a(X)}]/[{64X^4\phi^2(X)F_b(X)}],\nonumber
\label{eq:FG}
\end{equation}
\begin{equation}
\begin{split}
    G_a(X) &= 9\,b ( 4\,\ln( b)  ( -220\,{b}^{8}+456\,{b}^{6
}-303\,{b}^{4}+6\,\ln( b) ( 61\,{b}^{6}-69\,{b}^{4
}\\
&+4\,\ln( b)( 4\,\ln( b ) {b}^{4}-12
\,{b}^{4}+7\,{b}^{2}+2) {b}^{2}+9\,{b}^{2}+9) {b}^{2}+58
\,{b}^{2}+9) {b}^{2}\\
&+ ( {b}^{2}-1) ^{2}( 153
\,{b}^{6}-145\,{b}^{4}+53\,{b}^{2}-1)), \quad \mbox{where } b = 1+X \nonumber
\label{eq:Ga}
\end{split}
\end{equation}
\begin{equation}
\begin{split}
    F_a(X) &= -301b^8 + 622b^6-441b^4+4\log(b)\{197b^6-234b^4+6\log(b)\\
    &\times[16\log(b)b^4 - 36b^4+22b^2+3]b^2+78b^2+4\}b^2+130b^2-10,\nonumber
    \label{eq:Fa}
\end{split}
\end{equation}
\begin{equation}
    F_b(X) = 17b^6+12\log(b)[2\log(b)b^2-3b^2+2]b^4-30b^4+15b^2-2. \nonumber
    \label{eq:Fb}
\end{equation}

The surface tension plays both a stabilizing and a destabilizing role in the dynamics of flows down vertical fibres. This is characterized by the interaction between an azimuthal curvature term in $\mathcal{Z}(h)$ and the streamwise curvature terms $h_{xx}$ in \eqref{eq:q}.
Following the approach in \cite{ji2019dynamics}, we also introduce a film stabilization term $\Pi(h)$.
As a result, the functional $\mathcal{Z}$ in \eqref{eq:q} consists of a destabilizing azimuthal curvature term $\beta/(\alpha(1+\alpha h))$ and a film stabilization term $\Pi(h)$,
\begin{equation}
    \mathcal{Z}(h) = \frac{\beta}{\alpha(1+\alpha h)}+\Pi(h),\qquad \Pi(h) = -\frac{A}{h^3},
    \label{aziStab}
\end{equation}
where the scaling parameter $\beta = \alpha^2/\kappa^2$, and
$A>0$ is a stabilization parameter. The last term of \eqref{aziStab} takes the functional form of the long-range disjoining pressure of the well-known van der Waals model for wetting liquids. In lubrication theory, $A$ typically refers to a Hamaker constant that characterizes microscopic quantities (\cite{RevModPhys.57.827,reisfeld1992non}). Here we choose the value of $A$ based on a coating thickness $\epsp$ below which a thin uniform fluid layer on the fibre is stable (see Section \ref{sec:stability}).

The coupled system \eqref{eq:main} accounts for the surface tension, gravity, azimuthal instabilities and moderate inertial effects.
For $A = 0$, this model is consistent with the first-order weighted residual boundary layer model studied in \cite{ruyer2008modelling,duprat2009spatial} except that their model includes a fully nonlinear azimuthal curvature term $\mathcal{Z} = \mathcal{Z}_{FCM}$,
\begin{equation}
    \mathcal{Z}_{FCM}(h) = \frac{\beta}{\alpha(1+\alpha h)} + \frac{\alpha h_x^2}{2(1+\alpha h)}.
    \label{fullCUrvature}
\end{equation}
In the low Reynolds number limit $\delta \to 0$, or by setting $\partial q/\partial t \equiv 0$,  \eqref{eq:q} gives an expression for $q$ in terms of $h$. Substituting this expression into \eqref{eq:h} leads to a single lubrication equation for $h$ which is equivalent to the model \eqref{eq:fiber_CM} studied in \cite{ji2019dynamics, craster2006viscous},
\begin{equation}
\label{eq:fiber_CM}
\frac{\partial}{\partial t} \left(h+\frac{\alpha}{2}h^2\right) + \frac{\partial}{\partial x}\left[\mathcal{M}(h)\left(1-\frac{\partial }{\partial x}\left[\mathcal{Z}(h)-\frac{\partial^2 h}{\partial x^2}\right]\right)\right] = 0,
\end{equation}
where the $\mathcal{M}(h) = h^3\phi(\alpha h)/[3\phi(\alpha)]$ is the mobility function. The form of $\mathcal{M}(h)$ has also been generalized to include Navier slip boundary conditions in \cite{ji2019dynamics}.

\section{Results}
\subsection{Stability analysis and film stabilization mechanism}
\label{sec:stability}
Next we examine the linear stability of the model \eqref{eq:main} and derive the stabilization parameter $A$ in \eqref{aziStab}.
We perturb a uniform layer $h \equiv \bar{h}$ and its corresponding flux $q \equiv \bar{q}$,
\begin{equation}
    h = \bar{h}+\gamma H, \qquad q = \bar{q}+\gamma Q, \quad \mbox{where } \gamma \ll 1.
    \label{expansion}
\end{equation}
Substituting the expansion \eqref{expansion} into \eqref{eq:q} leads to the $O(1)$ equation,
$
    \bar{q} = {[\bar{h}^3\phi(\alpha \bar{h})]}/{[3\phi(\alpha)]}.
$
After obtaining a single equation for $H$ by eliminating $Q$ using the $O(1)$ equation, 
we apply the Fourier mode decomposition
$
    H = H_1\exp(\imagi k x + \Lambda t)
$,
where $k$ is the wavenumber and $\Lambda$ is the growth rate of the perturbation. This yields the dispersion relation,
    \begin{equation}
    \begin{split}
         \Lambda & = -\frac{3I(\alpha\bar{h})}{2\delta\bar{h}^2\phi(\alpha\bar{h})} + \frac{1}{2\delta}\sqrt{\frac{9I^2(\alpha\bar{h})}{\bar{h}^4\phi^2(\alpha\bar{h})} - \frac{4\delta}{1+\alpha\bar{h}}\left[-k^2\mathcal{S}(\bar{h},\bar{q})+\frac{I(\alpha \bar{h})\bar{h}}{\phi(\alpha)}k^4\right]},\\
       & \mbox{where}\quad \mathcal{S}(\bar{h},\bar{q}) = \delta G(\alpha \bar{h})\frac{\bar{q}^2}{\bar{h}^2}+\frac{I(\alpha \bar{h}) \bar{h}}{\phi(\alpha)}\left(\frac{\beta}{(1+\alpha \bar{h})^2}-\frac{3A}{\bar{h}^4}\right).
    \end{split} 
    \label{eq:dispersion}
\end{equation}
For the case $\mathcal{S} > 0$, we have $\Lambda > 0$ for $0 <k < k_c$, where the critical wavenumber $k_c = \sqrt{\mathcal{S}(\bar{h},\bar{q})\phi(\alpha)/[I(\alpha\bar{h})\bar{h}]}$. For the case $\mathcal{S} < 0$, we have $\Lambda < 0$ for any $k > 0$. 
We follow the approach in \cite{ji2019dynamics} and select the stabilization parameter $A$ based on the dimensional thickness $\epsp = \mathcal{H}h_c$ of a stable undisturbed layer obtained from experimental observations, where $h_c$ is the dimensionless stable coating film thickness. By setting $\mathcal{S}(h_c) = 0$ and using the $O(1)$ equation, we derive a formula for a critical $A_c$,
\begin{equation}
    A_c = \frac{\beta h_c^4}{3(1+\alpha h_c)^2}+\frac{\delta h_c^7G(\alpha h_c)\phi^2(\alpha h_c)}{27I(\alpha h_c)\phi(\alpha)}.
    \label{Aform}
\end{equation}
This form ensures that for $A=A_c$, any thin flat film of thickness less than the threshold value $h_c$ is linearly stable, that is, $\Lambda<0$, for all wavenumber $k > 0$.
Compared with the film stabilization model introduced in \cite{ji2019dynamics} (equivalent to the case $\delta = 0$), the formula \eqref{Aform} for $A_c$ includes a higher order term in $h_c$. 

\subsection{Near-nozzle flow dynamics}
\label{sec:numerics}
We perform numerical investigations to examine the spatio-temporal dynamics of the flow near the inlet and further downstream by solving the coupled system \eqref{eq:main} for $0 \le x \le L$.
To model the influence of nozzle geometry on the full dynamics, we impose Dirichlet boundary conditions on both the film thickness $h$ and the flux $q$ at $x = 0$: $ h(0, t) = \hin, q(0, t) = {1}/{3},$
where the dimensionless inlet film thickness $\hin = (\textstyle{\frac{1}{2}}\OD-R^*)/\mathcal{H}$ and $\OD$ represents the dimensional outer nozzle diameter. We select $\OD$ as the nozzle geometric parameter since the liquid wets the nozzle in experiments.
Following \cite{ruyer2008modelling}, we impose soft boundary conditions at the outlet $x = L$ by replacing the averaged momentum balance equation \eqref{eq:q} with a linear hyperbolic equation $q_t + v_f q_x = 0$ at the last two grid points near the outlet, using $v_f = 1$ in all simulations.
The initial conditions are a piece-wise linear profile for the film thickness $h$ and a constant for the flux $q$
\begin{equation}
q(x, 0) \equiv \frac{1}{3},\qquad
    h(x, 0) = 
    \begin{cases}
    1, \quad & x > x_L\\
    \hin+(1-\hin)x/x_L, \quad & 0\le x \le x_L
    \end{cases}
\end{equation}
where $x_L = 10$ is used for all simulations.

Centered finite differences in a Keller box scheme are used for numerically solving the model \eqref{eq:main}, where the coupled fourth-order PDE system is decomposed into a system of first-order differential equations:
\begin{equation}
\begin{split}
    &k = h_x,\quad p = \mathcal{Z}(h)-k_x, \quad (h+\textstyle\frac{\alpha}{2}{h^2})_t + q_x = 0,\\
   &\delta q_t =  \delta\left(- F(\alpha h)\frac{q}{h}q_x + G(\alpha h)\frac{q^2}{h^2}k\right)+\frac{I(\alpha h)}{\phi(\alpha)}\left[-\frac{3\phi(\alpha)}{\phi(\alpha h)}\frac{q}{h^2}+ h(1-
   p_x)\right].
   \label{eq:keller}
\end{split}
\end{equation}

\begin{figure}
    \centering
    \includegraphics[width=0.95\textwidth]{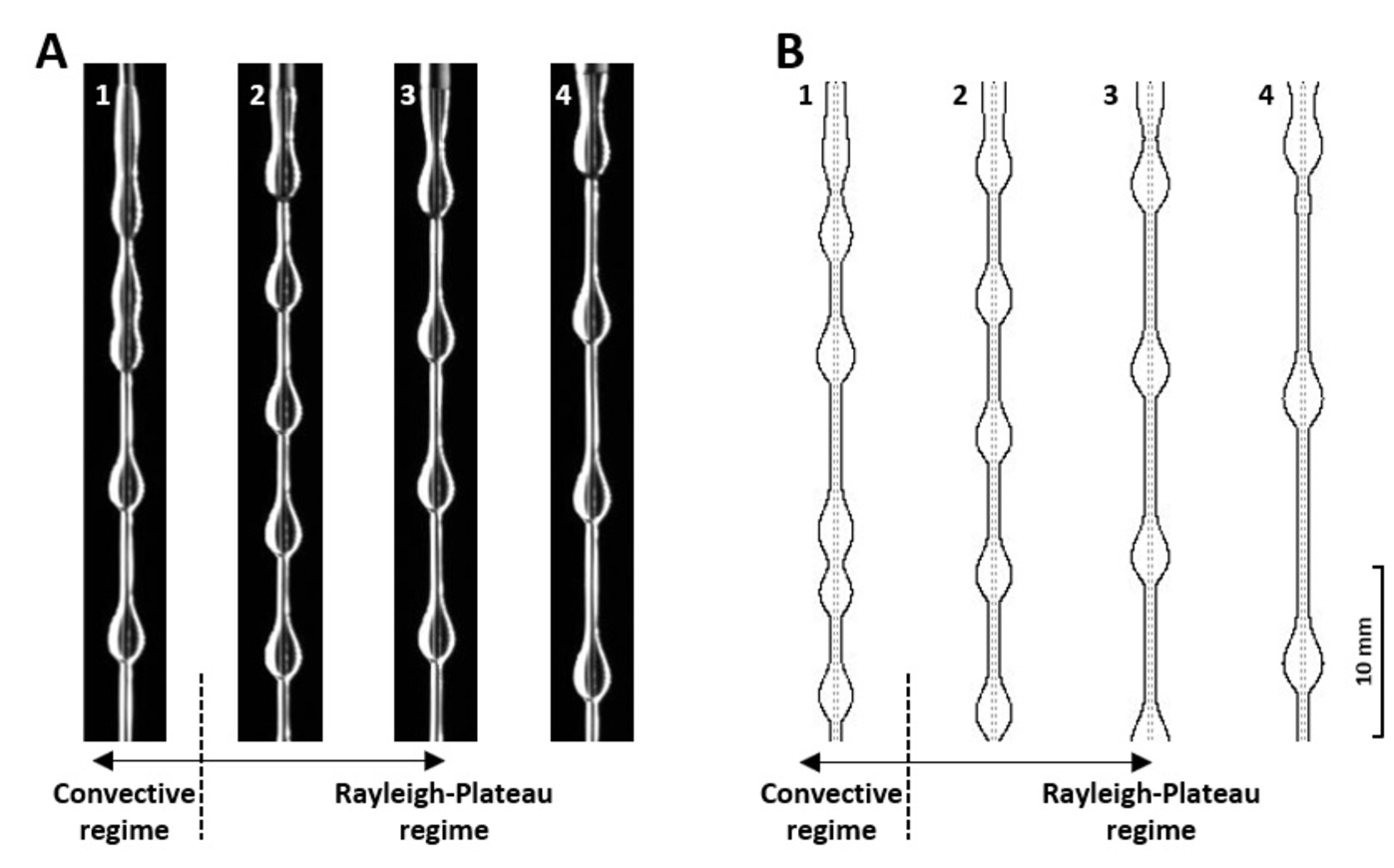}
    \caption{(A) Experiments and (B) numerical simulations of \eqref{eq:main} for nozzle diameters $\OD = 0.84, 1.27, 1.56, 1.86$ mm showing the transition from the convective to the Rayleigh-Plateau regime as $\OD$ increases from $\OD=0.84$ mm to $\OD=1.27$ mm. Other system parameters are $\Ds = 0.29$ mm, $\Qm = 0.04$ g/s, $\epsp = 0.15$ mm.}
    \label{fig:regime_transition}
\end{figure}

In figure~\ref{fig:regime_transition}, we show transient numerical results of model \eqref{eq:main} for four different nozzle diameters used in our experiments. A fixed flow rate $\Qm=0.04$ g/s and a fixed fibre diameter $\Ds=0.29$ mm are used. The experimental results in Fig.~\ref{fig:regime_transition} (A) indicate that as the nozzle diameter increases from $0.84$ mm to $1.27$ mm, the droplet dynamics undergo a  transition from the convective instability regime to the Rayleigh-Plateau regime. Moreover, within the Raleigh-Plateau regime, the larger nozzle diameter leads to larger spacing between the moving droplets. 
This regime transition is captured in the numerical simulation (see Fig.~\ref{fig:regime_transition} (B)) where the inter-bead spacing agrees well with experimental observations.

\begin{figure}
    \centering
    \includegraphics[width=0.95\textwidth]{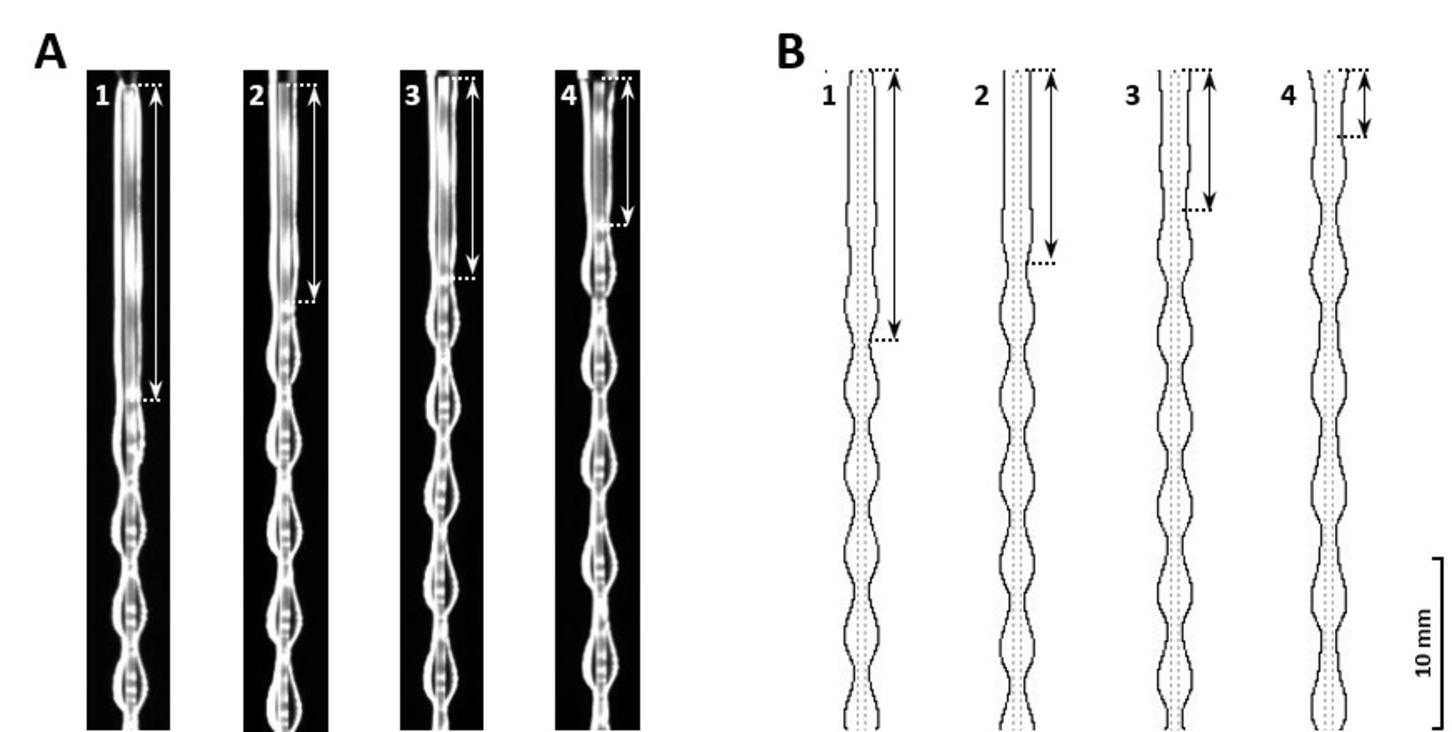}
    \caption{(A) Experiments and (B) numerical simulations with the nozzle diameter ranging from $\OD = 0.84$ to $2.41$ mm showing that the healing length (marked by arrows) decreases as the nozzle size increases. Other parameters are $\Ds = 0.43$ mm, $\Qm = 0.06$ g/s, $\epsp = 0.15$ mm.}
    \label{fig:healingLength}
\end{figure}

Next we discuss the influence of nozzle geometry on the flow dynamics within the healing length.
Previous studies assumed that the film thickness and velocity profiles in this part of the flow are specified only by the flow rate and fibre radius.
However, our study reveals that for flows in the convective and Rayleigh-Plateau regimes, the healing length decreases as the nozzle diameter increases.
Figure~\ref{fig:healingLength} presents one such comparison of the near-nozzle film profiles between experiments and simulations for the flow rate $\Qm=0.06$g/s and the fibre diameter $\Ds=0.43$ mm. This observation is reminiscent of the study by \cite{duprat2009spatial} on the spatial response of the film to inlet forcing concluding that the healing length tends to decrease as the forcing amplitude increases. 

\subsection{Liquid bead properties}
\label{sec:beads}

\begin{figure}
    \centering
    \includegraphics[width=0.97\textwidth]{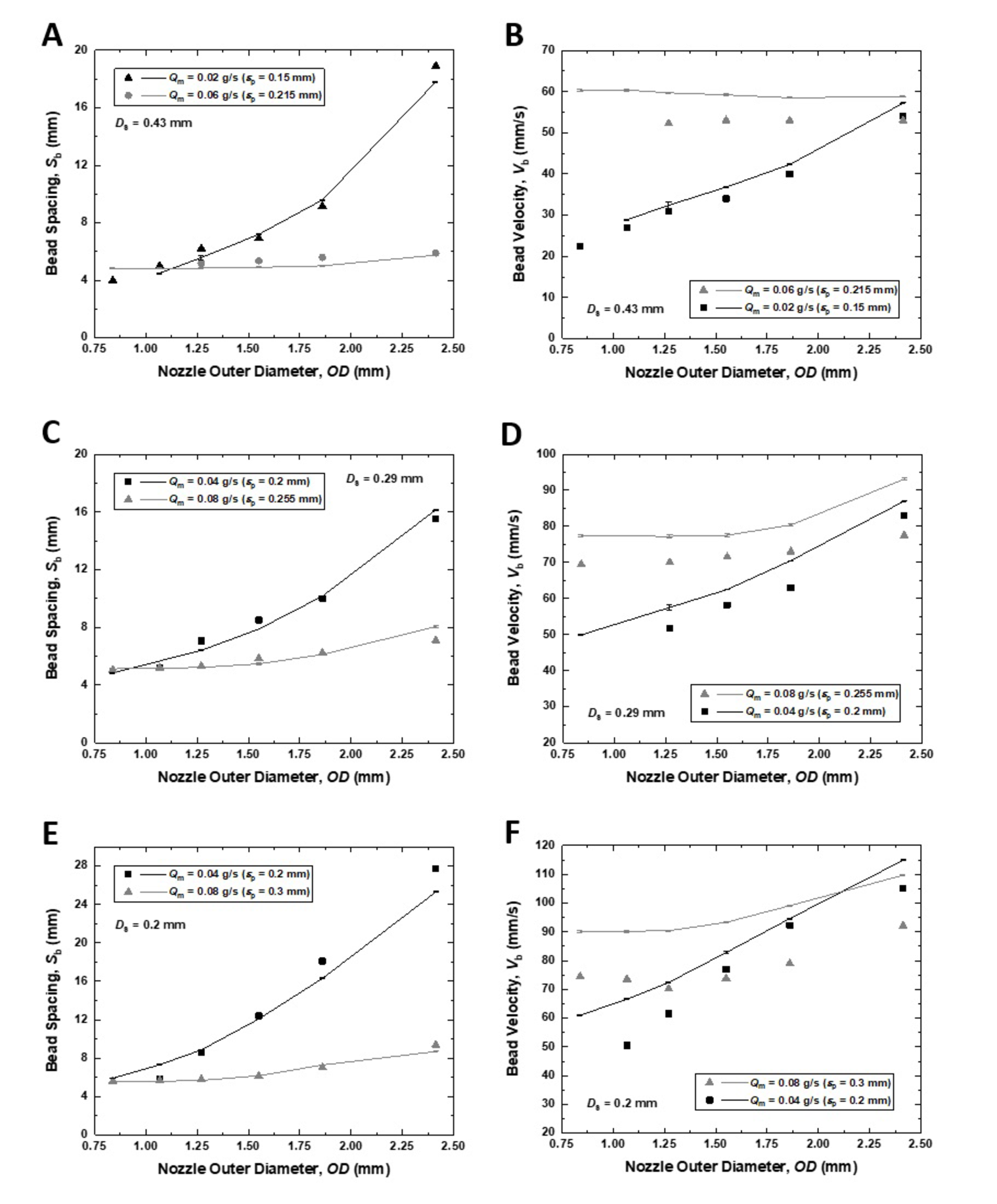}
    \caption{Bead properties (spacing and velocity) for the best $\epsp$ values. For thin fibres of diameter $\Ds=0.2$mm, we choose $\epsp = 0.2, 0.3$mm for $\Qm = 0.04, 0.06$ g/s, respectively. For the intermediate fibre  $\Ds=0.29$mm cases, we set $\epsp = 0.2, 0.255$mm for $\Qm = 0.04, 0.06$g/s. For the thick fibre  $\Ds=0.43$mm cases, we set $\epsp = 0.15, 0.215$mm for $\Qm = 0.02, 0.06$g/s, respectively. }
    \label{fig:beadProperties}
\end{figure}

In figure~\ref{fig:beadProperties}, we show plots of the predicted bead velocities $V_b$ and inter-bead spacing $S_b$ for varying nozzle sizes. The top through bottom panels show results with fibre diameters $\Ds=0.2, 0.29, 0.43$ mm respectively, with two choices of flow rates $\Qm$ for each fibre diameter. Our model \eqref{eq:main} agrees well with the experimental observations across all the cases, provided a suitable stable film thickness $\epsp$ is applied.  

\begin{figure}
    \centering
    \includegraphics[width=0.97\textwidth]{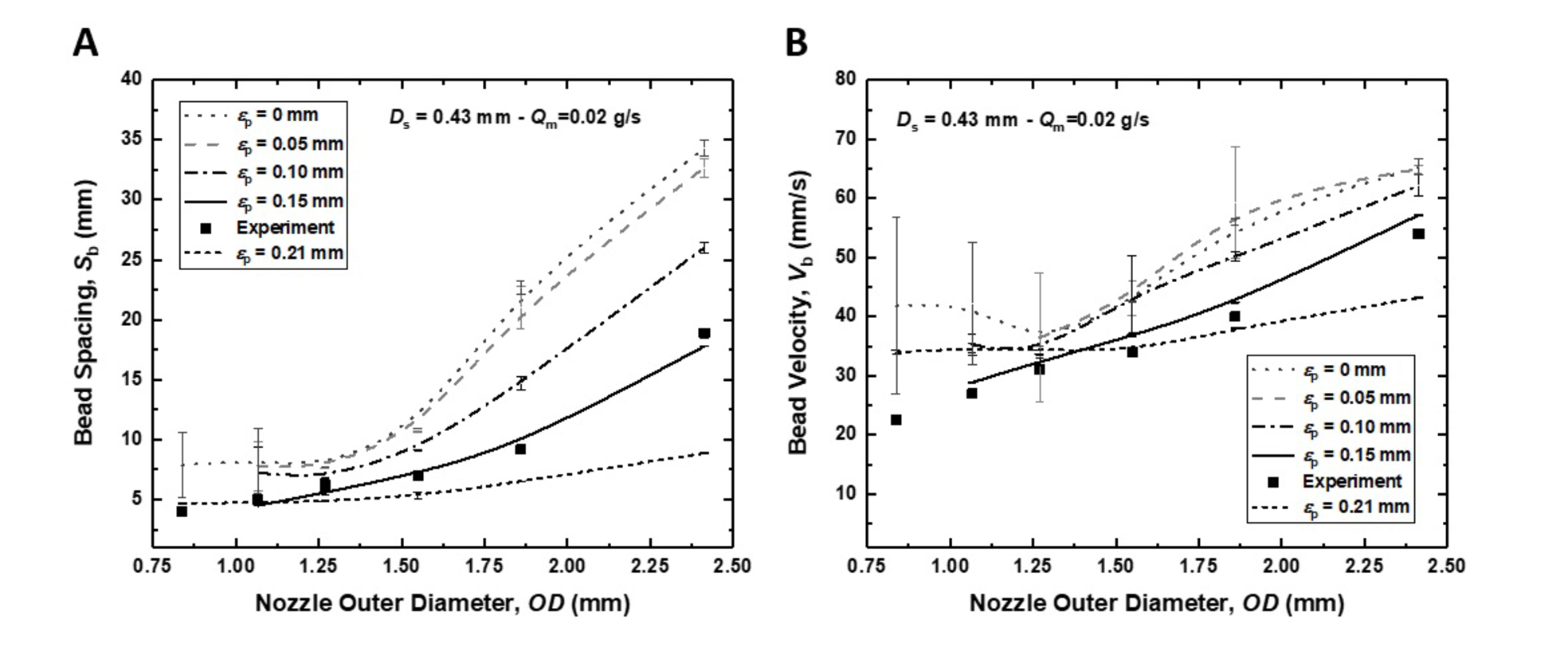}
    \caption{Averaged bead spacing and velocity obtained from experiments with $\Ds=0.43$~mm and $\Qm = 0.02$~g/s, compared to the characteristics predicted by the model \eqref{eq:main} for $A = A_c$ in \eqref{Aform} with a varying $\epsp = 0, 0.05, 0.1, 0.15, 0.21$mm.
}
    \label{fig:epsp}
\end{figure}

The presence of the film stabilization term is important for maintaining a stable train of beads flowing down the fibre. Figure~\ref{fig:epsp} shows the relation between the bead characteristics and the nozzle outer diameter $\OD$ for varying film stabilization thicknesses $\epsp$. Since a larger value of the stabilization parameter $A$ in \eqref{aziStab} corresponds to stronger stabilization effects in \eqref{eq:main}, increasing $\epsp$ is expected to yield more stabilized moving beads.
For a fibre diameter $\Ds=0.43$ mm at a small flow rate $\Qm=0.02$ g/s, the film stabilization model with $\epsp=0.15$ mm best captures both bead profiles and speeds. Without the film stabilization mechanism ($\epsp=0$), the model produces large variations in the downstream bead characteristics, contradicting experimental observations of a stable train of beads.

\begin{figure}
    \centering
    \includegraphics[width=0.97\textwidth]{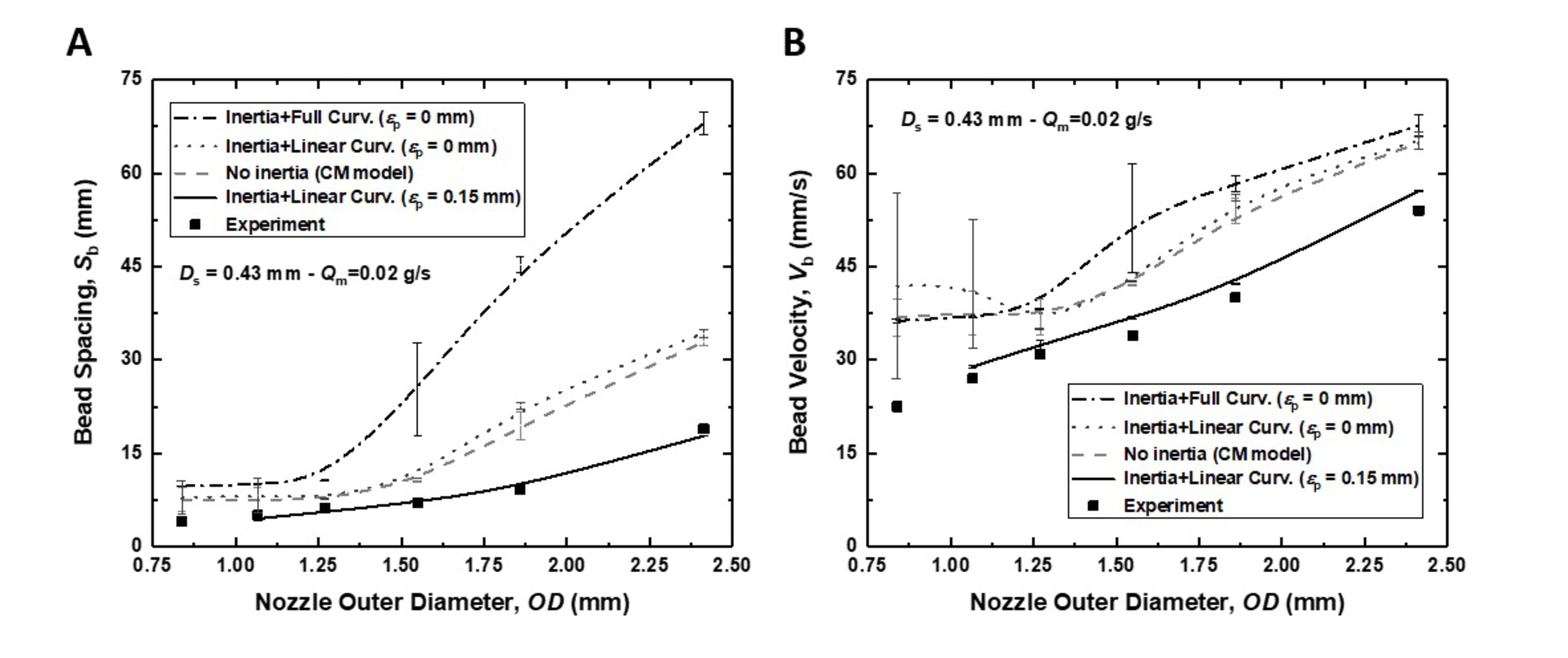}
    \caption{Averaged bead spacing and velocity obtained from experiments with $\Ds=0.43$~mm and $\Qm = 0.02$~g/s, compared to the characteristics predicted by the CM model \eqref{eq:fiber_CM} (without inertial effects), the full curvature model \eqref{fullCUrvature} and linear curvature model \eqref{aziStab} with $A = 0$ ($\epsp = 0$), and the film stabilization model \eqref{aziStab} with $A > 0$ ($\epsp = 0.15$~mm). 
}
    \label{fig:modelComparison}
\end{figure}

Lastly, we study the influence of inertial effects, nonlinear curvature terms, and the film stabilization term on the bead characteristics. Figure \ref{fig:modelComparison} shows a comparison of the experimental bead spacing and downstream bead velocity against those obtained from the Craster \& Matar model (CM) in \eqref{eq:fiber_CM}, the full curvature model with $\mathcal{Z}(h)$ given by \eqref{fullCUrvature} and $A = 0$, the linear curvature model with $\mathcal{Z}(h)$ in \eqref{aziStab} and $A = 0$, and the film stabilization model \eqref{aziStab} with $A > 0$. Whereas these models all yield qualitatively reasonable trends as the nozzle size increases, the film stabilization model provides the best quantitative agreement with the experiment. The other models over-estimate the bead spacing and velocity. For large nozzles ($\OD \ge 1.27$ mm), the CM model \eqref{eq:fiber_CM} with the film stabilization term for $\epsp = 0.15$ mm also predicts bead characteristics that match well with experiments (not shown in the figure). However, without inertial effects, this model fails to predict a steady train of beads for smaller nozzles ($\OD \le 1.06$ mm). 

\section{Conclusions}
\label{sec:conclusion}

We have performed a detailed study of viscous flows of a thin liquid film down vertical fibres, focusing on the influence of the inlet nozzle diameter on the regime transition and downstream bead dynamics. We propose a boundary layer model that incorporates a film stabilization term to the pressure, and compare the predicted film dynamics to a range of experiments. Numerical simulations show that, in addition to the fibre size and flow rate, the downstream flow regime transitions and bead characteristics are also affected by the nozzle geometry. In the Rayleigh-Plateau regime, our simulation results show good experimental agreement with the observed bead spacing and velocity throughout the fibre and with the film profile within the healing length near the nozzle.

% \section{Acknowledgements}
% Acknowledgements should be included at the end of the paper, before the References section or any appendicies, and should be a separate paragraph without a heading. 
This work was supported by the Simons Foundation Math+X investigator award number 510776 and the National Science Foundation under grant CBET-1358034.

\vspace{0.1in}
\noindent
\emph{Declaration of Interests:}
The authors report no conflict of interest.

% \newpage
%\nocite{*}
\bibliographystyle{jfm}
\bibliography{NozzleFiberFilms}
\end{document}